\input harvmac
\input epsf
\def\figin{\epsfcheck\figin}\def\figins{\epsfcheck\figins}
\def\epsfcheck{\ifx\epsfbox\UnDeFiNeD
\message{(NO epsf.tex, FIGURES WILL BE IGNORED)}
\gdef\figin##1{\vskip2in}\gdef\figins##1{\hskip.5in}
\else\message{(FIGURES WILL BE INCLUDED)}%
\gdef\figin##1{##1}\gdef\figins##1{##1}\fi}
\def\DefWarn#1{}
\def\figinsert{\goodbreak\midinsert}
\def\ifig#1#2#3{\DefWarn#1\xdef#1{fig.~\the\figno}
\writedef{#1\leftbracket fig.\noexpand~\the\figno}%
\figinsert\figin{\centerline{#3}}\medskip\centerline{\vbox{\baselineskip12pt
\advance\hsize by -1truein\noindent\footnotefont{\bf Fig.~\the\figno:} #2}}
\bigskip\endinsert\global\advance\figno by1}

\def\r{R_{11}}
\def\x{x^{11}}

\def\k{N_c}
\def\N{\nu}
\def\gluino{gaugino}
\def\gluinos{gauginos}
\def\SYM{super Yang-Mills}

\def\mqcd{\hsz\berkM\ozdb\kapl\berkMM}

\def\bb#1{ hep-th/#1}

\lref\vafa{C. Vafa, {\it Gas of D-Branes and Hagedorn Density of BPS States},
hep-th/9511088, Nucl.Phys. B463 (1996) 415-419.}

\lref\herman{F. Hacquebord, H. Verlinde, 
{\it Duality symmetry of N=4 Yang-Mills theory on $T^3$.}, hep-th/9707179,
Nucl.Phys. B508 (1997) 609-622.}

\lref\hw{A. Hanany and E. Witten, {\it Type IIB Superstrings, 
BPS Monopoles, And
Three-Dimensional Gauge Dynamics.}, 
hep-th/9611230, Nucl.Phys. B492 (1997) 152-190.}

\lref\merons{A. Actor, {\it Classical Solutions of 
$SU(2)$ Yang-Mills Theory}, Rev.Mod.Phys. 51, 461(1979).}

\lref\cdg{C. Callan, Dashen, D. Gross, {\it Toward a Theory of the 
Strong Interactions}, Phys.Rev.D17:2717,1978.}

\lref\dvv{R. Dijkgraaf, E. Verlinde, and H. Verlinde, {\it 
BPS Spectrum of the Five-Brane and Black Hole Entropy,} 
hep-th/9603126,
Nucl.Phys. B486 (1997) 77-88; {\it BPS Quantization of the Five-Brane},
hep-th/9604055, Nucl.Phys. B486 (1997) 89-113.}

\lref\vw{Cumrun Vafa, Edward Witten, 
{\it A Strong Coupling Test of S-Duality},hep-th/9408074, 
Nucl. Phys. B431 (1994) 3-77.}

\lref\strominger{A. Strominger, {\it Open P-Branes}, 
hep-th/9512059, Phys.Lett. B383 (1996) 44-47.}

\lref\douglas{ M.R. Douglas, {\it Branes within Branes},  RU-95-92,
hep-th/9512077.}

\lref\smallinst{E. Witten, {\it Small Instantons in String Theory,}
Nucl.Phys. B460 (1996) 541-559, hep-th/9511030.}

\lref\qcd{E. Witten {\it Branes And The Dynamics Of QCD}, 
Nucl.Phys. B507 (1997) 658-690.} 

\lref\hsz{A. Hanany, M. J. Strassler, and A. Zaffaroni
{\it Confinement and Strings in MQCD},  IASSNS--HEP--97/91,
hep-th/9707244.}

\lref\ozdb{J. de Boer and Y. Oz,
{\it `` Monopole Condensation and Confining Phase of N=1 Gauge
Theories via M theory Five-Brane'',}\bb{9708044.}}

\lref\berkM{K. Hori, H. Ooguri and Y. Oz, {\it ``Strong Coupling
Dynamics of Four-Dimensional $N=1$ Gauge Theories from M Theory
Fivebrane",} \bb{9706082.}}

\lref\kapl{A. Brandhuber, N. Itzhaki, V. Kaplunovsky, J. Sonnenschein
and S. Yankielowicz, {\it ``Comments on the M Theory Approach to
$N=1$ $SQCD$ and Brane Dynamics",} \bb{9706127.}} 

\lref\berkMM{Jan de Boer, Kentaro Hori, Hirosi Ooguri, Yaron Oz, 
{\it Kahler Potential and Higher Derivative Terms from M
Theory Fivebrane}, hep-th/9711143.}

\lref\cumrun{C. Vafa, {\it Instantons on D-branes},
hep-th/9512078, Nuclear Physics B463 (1996) p.435.}

\lref\instcal{V.A. Novikov, M.A. Shifman, A.I. Vainshtein, V.I. Zakharov, 
{\it Instanton Effects in Supersymmetric Theories}, 
Nucl.Phys.B229:407,1983.}

\lref\shifman{A. Kovner, M. Shifman, A. Smilga, 
{\it Domain Walls in Supersymmetric Yang-Mills Theories},
hep-th/9706089, Phys.Rev. D56 (1997) 7978-7989.}

\lref\amiofer{Ofer Aharony, Amihay Hanany, 
{\it Branes, Superpotentials and Superconformal Fixed Points}, 
hep-th/9704170, Nucl.Phys. B504 (1997) 239-271; 
S. Elitzur, A. Giveon, D. Kutasov, E. Rabinovici, A. Schwimmer, {\it 
Brane Dynamics and N=1 Supersymmetric Gauge Theory,}
hep-th/9704104,
Nucl.Phys. B505 (1997) 202-250.}

\lref\bending{E. Witten, {\it Solutions Of Four-Dimensional Field 
Theories Via M Theory}, Nucl.Phys. B500 (1997) 3-42.}

\lref\torons{G. 't Hooft, {\it Some Twisted Self-dual Solutions for the 
Yang-Mills Equations on a Hypertorus}, Commun.Math.Phys.81:267-275,1981.}

\lref\matrixreview{T. Banks, L. Susskind}
\lref\comments{E. Witten, ``Some Comments on String Dynamics'' }

\lref\coleman{S. Coleman, {\it Aspects of Symmetry}, 
Cambridge University Press, 1985.}

\lref\karch{SIX-DIMENSIONAL THEORIES WITH BRANES}

\lref\watiREVIEW{W. Taylor, {\it 
Lectures on D-branes, Gauge Theory and M(atrices)},
PUPT-1762, hep-th/9801182.}

\lref\neqoneREVIEW{K. Intriligator and N. Seiberg, {\it 
Lectures on supersymmetric gauge theories and
electric-magnetic duality}, hep-th/9509066, 
Nucl.Phys.Proc.Suppl. 45BC (1996) 1-28; 
Nucl.Phys.Proc.Suppl. 55B (1996) 200-209 ; M. Peskin, 
{\it Duality in Supersymmetric Yang-Mills Theory}, 
SLAC-PUB-7393, hep-th/9702094, TASI 96 Proceedings.}

\lref\kenscott{K. Intriligator and S. Thomas, 
{\it Dynamical Supersymmetry Breaking on Quantum Moduli
Spaces.}, hep-th/9603158, Nucl.Phys. B473 (1996) 121-142.}

\lref\iy{ Izawa K.-I., T. Yanagida, 
{\it Dynamical Supersymmetry Breaking in Vector-like Gauge Theories,} 
Prog.Theor.Phys. 95 (1996) 829-830, hep-th/9602180.}

\lref\berkeleyIT{Jan de Boer, Kentaro Hori, Hirosi Ooguri, Yaron Oz, {\it 
Branes and Dynamical Supersymmetry Breaking}, hep-th/9801060.}

\lref\branesREVIEW{A. Giveon and D. Kutasov, {\it 
Brane Dynamics and Gauge Theory}, RI-2-98, EFI-98-06, hep-th/9802067.}

\lref\barbon{J.L.F. Barbon, A. Pasquinucci, {\it D0-Branes as Instantons in 
D=4 Super Yang-Mills Theories}, CERN-TH/97-354, KUL-TF-97/35, hep-th/9712135.}

\lref\barbonTWO{J.L.F. Barbon, A. Pasquinucci, {\it D0-Branes, Constrained Instantons and D=4 Super
Yang-Mills Theories}, CERN-TH/97-196, IFUM-576/FT, hep-th/9708041.}

\lref\keg{S. Elitzur, A. Giveon, D. Kutasov, {\it Branes and 
N=1 Duality in String Theory}, hep-th/9702014, Phys.Lett. B400 (1997) 269-274.}

\lref\ahiss{O. Aharony, A. Hanany, K. Intriligator, M. Strassler, N. Seiberg, 
{\it Aspects of N=2 Supersymmetric Gauge Theories in Three
Dimensions.} hep-th/9703110, Nucl.Phys. B499 (1997) 67-99.}

\lref\cho{P. Cho and P. Kraus, {\it Symplectic SUSY Gauge Theories with Antisymmetric
Matter,} Phys.Rev. D54 (1996) 7640-7649, hep-th/9607200.}

\lref\skiba{Csaba Csaki, Witold Skiba (MIT), Martin Schmaltz, {\it 
Exact Results and Duality for SP(2N) SUSY Gauge Theories
with an Antisymmetric Tensor,} Nucl.Phys. B487 (1997) 128-140, hep-th/9607210;
G. Dotti and A. Manohar, {\it Supersymmetric Gauge Theories with an Affine Quantum
Moduli Space,} hep-th/9712010,  Phys.Rev.Lett. 80 (1998) 2758-2761.}

\lref\bci{J. Brodie, P. Cho, and K. Intriligator, {\it 
Misleading Anomaly Matchings?,} HUTP-98/A001, PUPT-1758, UCSD/PTH 98-04, 
hep-th/9802092.}

\lref\juan{J. Maldacena, {\it Statistical Entropy of Near Extremal Five-branes,} Nucl.Phys. B477 (1996) 168-174, hep-th/9605016.}

\lref\aki{A. Hashimoto, {\it Perturbative Dynamics of Fractional Strings on Multiply
Wound D-strings,} PUPT-1657, hep-th/9610250.}

\lref\rajesh{Rajesh Gopakumar {\it BPS states in Matrix Strings,} 
Nucl.Phys. B507 (1997) 609-620, hep-th/9704030.}

\lref\gavin{Gavin Polhemus, 
{\it Statistical Mechanics of Multiply Wound D-Branes,}
Phys.Rev. D56 (1997) 2202-2205, hep-th/9612130.}

\lref\yi{K. Lee and P. Yi, {\it Monopoles and Instantons on Partially Compactified
D-Branes,} hep-th/9702107, Phys.Rev. D56 (1997) 3711-3717; 
K. Lee {\it Instantons and Magnetic Monopoles on $R^3\times S^1$ with
              Arbitrary Simple Gauge Groups,} hep-th/9802012; 
C. Vafa, {\it 
On N=1 Yang-Mills in Four Dimensions,} hep-th/9801139.}
\lref\brodieami{J. Brodie and A. Hanany, {\it Type IIA Superstrings, Chiral Symmetry, and N=1 4D Gauge
Theory Dualities,} Nucl.Phys. B506 (1997) 157-182, hep-th/9704043.}

\lref\braneswithin{M. Douglas, {\it Branes within branes,}.}


\def\LongTitle#1#2#3#4#5{\nopagenumbers\abstractfont
\hsize=\hstitle\rightline{#1}
\hsize=\hstitle\rightline{#2}
\hsize=\hstitle\rightline{#3}
\vskip 0.5in\centerline{\titlefont #4} \centerline{\titlefont #5}
\abstractfont\vskip .3in\pageno=0}

\LongTitle{PUPT-1774}{hep-th/yyymmnn}{}
{Fractional Branes, Confinement, and} 
{Dynamically Generated Superpotentials}

\centerline{
  John Brodie}
\bigskip
\centerline{ Department of Physics, Princeton University,
  Princeton, NJ 08540}

\vskip 0.3in
\centerline{\bf Abstract}
\bigskip

We examine the effects of instantons in four-dimensional 
$N=1$ supersymmetric 
gauge theory by including D0-branes in type IIA brane 
constructions.
We examine instanton-generated superpotentials 
in supersymmetric QCD and find that they are due to a 
repulsive force between D4-branes bound to D0-branes ending on NS 5-branes. 
We study situations 
where instanton effects break supersymmetry such as 
the Intriligator-Thomas-Izawa-Yangagida model and relate this 
to a IIA brane construction.
We also argue how 
confinement due to a condensate of fractional instantons 
manifests itself in Super Yang-Mills theory using fractional
D0 branes, D4 branes, and NS strings.

\Date{1/98}


\newsec{Introduction}

There has been much progress in understanding 
supersymmetric gauge theories using 
brane constructions
\branesREVIEW. 
Generalizing a 
three-dimensional brane construction of 
\hw, the authors of \keg\ were able to 
reproduce Seiberg's 
$N=1$ supersymmetric gauge theory duality 
\neqoneREVIEW\ using 
brane constructions in type IIA string theory.
In \bending, it was shown how to obtain
the Seiberg-Witten curve 
describing the exact quantum Coulomb branch of 
$N=2$ gauge theories. 
A configuration of D4-branes and  
NS 5-branes in type IIA string theory preserving $1/4$ supersymmetry
gives the classical $N=2$ Super Yang-Mills 
worldvolume field theory.
Lifting the IIA configuration to eleven-dimensional M-theory,
the brane construction becomes a single 5-brane and
the four-dimensional 
supersymmetric gauge theory on the worldvolume 
of the D-branes
decompactifies into a six dimensional 
self-dual tensor theory. 
Remarkably, the M-theory limit 
provides quantum information about the 
four-dimensional
field theory on the worldvolume of the IIA branes.
Results about $N=1$ Super Yang-Mills were also 
obtained 
by considering configurations of 
D4-branes and NS 5-branes, preserving only 
$1/8$ the supersymmetry of type IIA string theory.
Strings and domain walls
were
obtained by relating the IIA configuration 
a single M-theory 5-brane \qcd. 
The topic of obtaining field theory 
results about $N=1$ 
supersymmetric gauge theory from M-theory 
is known as MQCD and   
was pursued in a number of directions
\mqcd.
In this paper we will investigate 
quantum effects in 
worldvolume gauge theories 
by another route. We will construct brane configurations 
in IIA string theory with 4 supercharges, but instead 
of lifting the configuration to M-theory, 
we will remain in type IIA string theory, keeping 
the gauge theory four-dimensional. 
To obtain quantum results about the 
worldvolume gauge theory, 
we will introduce D0 branes into the D4-brane worldvolume 
as was done in \barbon\barbonTWO. 
Because a D0 brane inside 
a D4 brane is a Yang-Mills instanton \smallinst\douglas, 
including D0 branes in the 
string theory brane constructions should be 
equivalent to including instanton effects 
in the worldvolume gauge theory.
D0 branes are Kaluza-Klein modes 
of the compactified eleventh dimension of M-theory, and 
therefore
our approach should in some sense be the ``Fourier transform''
of the MQCD approach. 

In section 2, we 
argue that fractional instantons are responsible for 
gaugino condensation and the mass gap in low energy 
$N=1$ $SU(N_c)$ \SYM\ theory by showing how fractional D0-branes 
induce a non-zero vacuum energy in the worldvolume theory on the 
branes.
We go on in section 3
to examine the brane construction for supersymmetric 
$SU(N_c)$ QCD with $N_f$ flavors 
and 
show how D0 branes generate the Afflect-Dine-Seiberg superpotential 
in the worldvolume theories lifting the degeneracy of vacua for the 
case of $N_f = N_c -1$ flavors. In 
brane language 
the superpotential manifests itself as a repulsive force between the 
D0-D4 brane bound state and other D4-branes ending on an NS 5-brane.
For the case of $N_f = N_c$ flavors, 
instantons can be seen to be responsible for the deformed 
moduli space using D0-branes in the brane constructions. 
In section 4, 
we show how instanton effects break supersymmetry in the 
Intriligator-Thomas model using brane diagrams: The non-zero vacuum 
comes from a repulsive force between D4-branes as in the 
superpotential case. The difference here is that 
the D4-branes are held together by heavier 
NS 5-branes and cannot run away to a zero energy configuration.
Therefore, there is a non-zero vacuum energy and supersymmetry 
must break.
In section 5, 
we examine the one instanton effect in $N=2$ \SYM\ theories and 
explain why fractional instantons do not play a role in 
the Seiberg-Witten solution. Introducing an orientifold plane, in section 6, 
takes 
the $N=2$ brane construction to $N=1$ $Sp(N_c)$ gauge theory 
with an antisymmetric tensor field. The vacuum of this theory 
has two branches: one with a dynamically generated superpotential and 
one that is the same as the classical vacua. 
We explain the dynamically generated 
superpotential as occurring in the brane construction where the
mass gap induced by 
Euclidean D0-brane charge,
momentum across a surface in the eleventh dimension, 
does not cancel and the branch in moduli space 
with no superpotential as coming 
from the brane configuration where the mass gap induced by the 
D0-branes exactly cancels.
In section 7,
we will consider a fundamental type IIA string 
inside a D4 brane. Such a configuration is only BPS if the 
string spreads out inside the worldvolume of the D4-brane, 
becoming unconfined electric flux in the 4+1 \SYM. 
By building a string from fractional Euclidean D0 branes, 
we will show that 
the string's lowest energy configuration no longer 
requires it to spread out.
Free electric flux therefore becomes confined to a 
tube. This appears to realize confinement by instantons.
We will apply this result to 
the brane construction of
$N=1$ $SU(N_c)$ \SYM\ and argue that fundamental strings 
bound to fractional Euclidean 
D0-branes behave like confined flux tubes.

\newsec{Gaugino condensation in $N=1$ \SYM.}
\seclab{\gauginoNS}

Here we construct $N=1$ $SU(\k)$ \SYM\ from branes. 
Consider a NS 5-brane extending in 
directions $(x^0,x^1,x^2,x^3,x^4,x^5)$ and a 
NS' 5-brane extending in directions 
$(x^0,x^1,x^2,x^3,x^8,x^9)$. $\k$ D4-branes 
extend in the directions 
$(x^0,x^1,x^2,x^3,x^6)$ and
end on the NS and the NS' 5-branes
in the direction $x^6$.
The coupling constant 
of the 3+1 $SU(k)$ $N=1$ \SYM\ theory living 
in $(x^0,x^1,x^2,x^3)$ is 
\eqn\coupling{{1\over g_3^2} = {L_6M_s\over g_s} = {L_6\over \r}.}
Euclidean D0 branes
extending in the $x^6$ direction appear 
as instantons in the four-dimensional gauge theory. 
This was noticed in \smallinst\douglas.
The action for the D4-brane worldvolume theory 
with D0-branes is 
\eqn\action{ {\cal L} = F^{(2)} \wedge *F^{(3)} 
+ A^{(1)}_{RR} \wedge F^{(2)} \wedge F^{(2)}}
where $A_{RR}^{(1)}$ is the Ramond-Ramond 1-form of 
type IIA string theory that couples to the D0 brane in ten 
dimensions and $F$ is the field strength of the \SYM\ 
gauge field on the 
D4 brane world volume.
Dimensionally reducing the D4-brane to four dimensions along $L_6$, 
instantons 
are self-dual solutions of the Yang-Mills field strength, 
\eqn\selfdual{F^{(2)} = *F^{(2)}.}
Using the fact that 
the Bianchi identity is identically satisfied $dF^{(2)} = 0$, 
instantons also satisfy the equations of motion of 
the Yang-Mills action $d*F^{(2)} = 0$. 
The action of the instanton is given by 
equation \coupling.
The $1/\r$ in \coupling\ we
expect since the D0 brane is 
momentum in the 
$\x$ direction across a surface in the $x^6$ direction. 
If we lift the D0-D4 system to M-theory we see that the 
energy-momentum tensor of the 
$(0,2)$ theory is given by
\eqn\T{T^{6,11} = H^{6,\mu.\nu}H^{\mu,\nu,11}.}
In the far IR, the gauginos condense giving rise to a mass gap.
The expectation value of the 
gaugino condensate goes as 
\eqn\gaugino{<\lambda \lambda> = \Lambda^3 = e^{-8\pi^2/N_cg_3^2}.}
From the exponent in \gaugino,
this non-perturbative effect has the interpretation as 
the effect of one fractional instanton.
Indeed, $N=1$ \SYM\ has $2N_c$ fermionic zero modes coming 
from the gauginos. Performing an instanton calculation using the 
't Hooft vertex, one can show that \instcal\ 
\eqn\Npoint{<\lambda\lambda(x_1) \lambda\lambda(x_2) ... 
\lambda\lambda(x_{N_c})> = \Lambda^{2N_c}.}
Cluster decomposition then implies that \Npoint\ 
decomposed into \gaugino.
However, instantons by themselves cannot 
produce \gaugino, whereas fractional instantons 
have just the right amount of zero modes.
Therefore, it is natural to expect that 
fractional instantons are responsible for the 
gaugino condensate \shifman.

\ifig\DoneDthree{On the right hand side we see $\k$ D4 branes with 
$\N$ D0 branes stuck to them. On the left we see the 
T-dual version where the D1 branes are free to 
split on the D3 branes. The split D1 branes are 
T-dual to the fractional instantons.}
{\epsfxsize3.0in\epsfbox{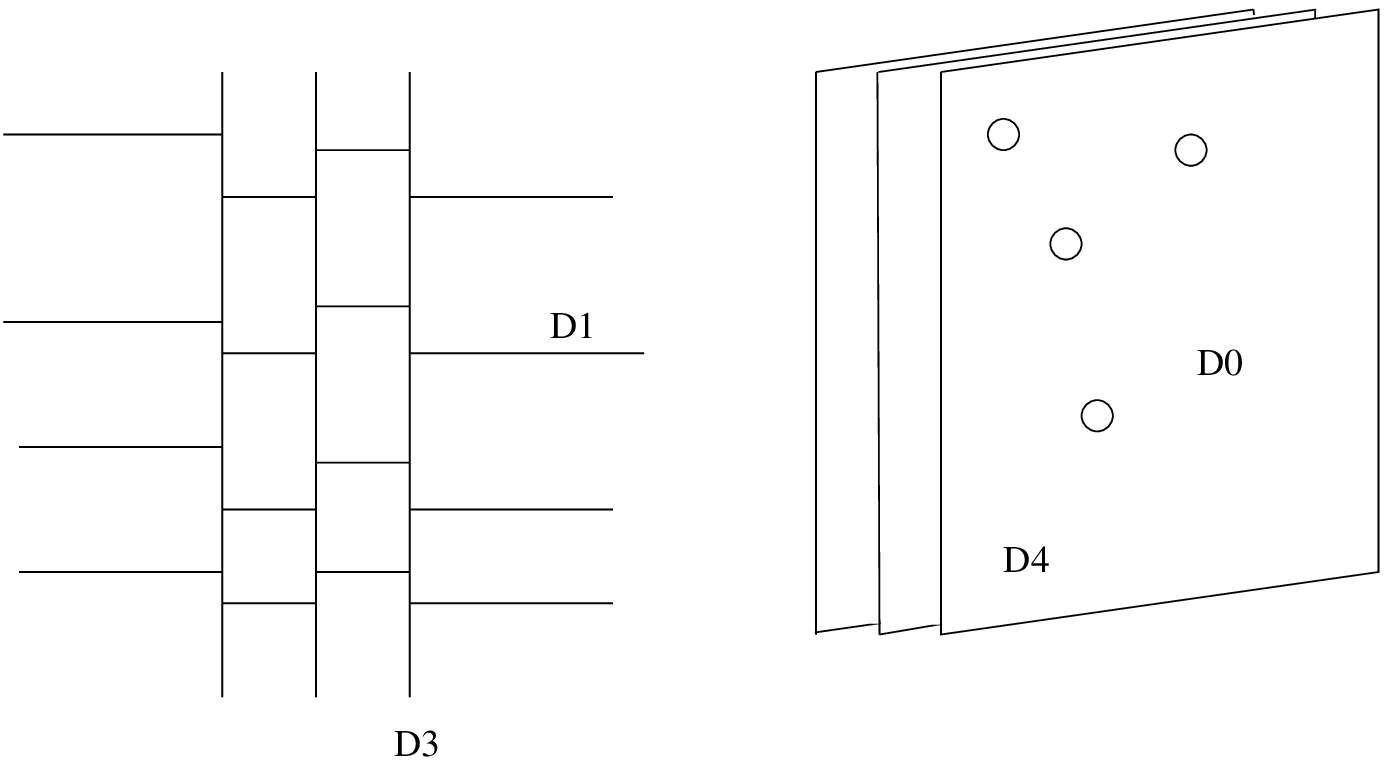}}
\subsec{Dimensional reduction of four-dimensional $N=1$ \SYM\ 
to three-dimensional $N=2$ \SYM.}
\subseclab{\threedimSS}

To understand equation \gaugino\ better, 
let us now consider compactification of the 
four-dimensional $N=1$ \SYM\ theory to three dimensions. 
In the brane construction this is done by compactifying 
the $x^3$ direction 
and T-dualizing along it yielding a IIB brane configuration. 
D4-branes become D3-branes and D0-branes 
become D1-branes. 
It is now possible to break the D1-branes on the D3-branes 
such that the D1-branes move between the D3-branes 
in the $x^0,x^1,x^2$ directions (see \DoneDthree). 
The points where the D1-branes meets the D3-branes looks like a monopoles 
in the worldvolume theory of the 
D3-branes. We can consider 
T-dualizing the D1-D3 system back to the D0-D4 system. 
What becomes of the broken D1-branes? 
We will argue here that 
they can be thought of as {\it fractional} D0-branes with charge 
$1/\k\r$ \merons \torons. In the 
D4-brane worldvolume these are precisely the fractional instantons that we 
need to explain \gaugino. 
The gauge potential for a meron, a fractional instanton on 
${\bf R^4}$, is 
\eqn\A{A_{\mu} = {1\over \k} g \partial_{\mu} g^{-1}.}
Notice that because of the ${1\over\k}$ out front this 
is not gauge equivalent to the vacuum.

For an $SU(\k)$ gauge theory on ${\bf R^4}$ the $\N$ 
instanton moduli space has dimension $4\N\k$. 
The dimension of instanton moduli space has the interpretation 
roughly as $\N$ instantons 
that in addition to $4\N$ translational zero modes have modes 
which correspond to scale size.
However, when the instantons becomes point-like they can fractionate. 
The fractional instanton moduli space is then 
a space of $\N\k$ fractional instantons
that can move on ${\bf R^4}$ in four different ways.
Instead of the dimension of the moduli space being associated with 
translations and ``fatness'', the dimension of 
fractional instanton moduli space is only associated with 
translations of point-like objects, as the T-dual D1-D3 system suggests.
This fits nicely with the fact that the cohomology of 
$\N$ $SU(\k)$ instantons on ${\bf T^4}$ is 
the same as $(T^4)^{\N\k}/S_{\N\k}$ implying  
$\N\k$ points on $T^4$ \vafa.
\foot{
Merons are point-like objects and have infinite action, 
so it appears at first sight that they are irrelevant to the partition 
function.
\eqn\action{e^{-S_I} = e^{-\infty} = 0.}
However, this is only true in the semi-classical approximation 
where we expand about 
the center-point of a Gaussian integral \coleman.}

In the $N=2$ three-dimensional $SU(N_c)$ brane construction,
$N_c$ D3-branes can separate in the 
direction $x^3$ breaking the 
$SU(N_c)$ gauge theory 
to $U(1)^{N_c}$. This corresponds to 
an vacuum expectation value for the 
real adjoint scalar field.
Although globally the 
supersymmetry is broken the four real supercharges, in actuality
locally
where the 
D3-branes end on the NS 5-brane, 
eight real supercharges are preserved.
Including D1-branes wrapping the 
$x^3$ direction, breaks the 
supersymmetry locally where the 
D3-brane meets the NS 5-brane to four real 
supercharges. 
Monopole-monopole interactions cancel only 
in theories with eight supersymmetries,
and therefore the D3-branes must repel each other.
If we take the $x^3$ direction to be compact, then 
the D3-brane will move around the circle to 
equally separated points. This was discussed in 
\amiofer\ and is consistent with field theory \ahiss.
Because the D3-branes are equally distributed around the circle 
this constrains the D1-branes to break up into equal sized pieces.
This is consistent with the fact that solutions of the Yang-Mills 
equations with fractional topological charge must come in units 
of $1/{N_c}$ rather than some arbitrary charge as can be seen in \A.
The splitting of the D1-branes on the 
D3-branes Higgses the 
$U(1)^{N_c}$ gauge group 
that is present on the 
world volume of the D3-branes.
We can see that 
there will be two D1-D3 strings 
on each fractionated D1-brane. 
There will be two fermion zero modes
coming from these strings. This 
is very reminiscent of 
\gaugino. Similar observations have been made in \yi.

\subsec{Return to four-dimensional $N=1$ \SYM.}
In the T-dual four-dimensional brane construction,
the D4-branes look like 
vortices in the NS 5-brane worldvolume. 
We expect the vortices to repel each other as 
do the monopoles in the three-dimensional case.
The mass gap comes about because
the D0-brane breaks the supersymmetry of the 
D4-branes to $1/8$ at the point where the D0-D4 bound state intersects the 
NS' 5-brane. Like the monopole-monopole interactions in 
the three-dimensional theory, the vortex-vortex interaction 
cancel in theories with eight supersymmetries. Therefore we 
expect for $N_c$ D4-branes in the presence of a 
D0 brane, there will be a repulsive force between the D4-branes.
However, because the NS and NS' 5-brane are holding 
the D4-branes in place, the D4-branes have no where to 
go. This force between the D4-branes produces 
a mass gap in the theory, consistent with field theory.
It would be interesting to understand the origin of 
the vortex-vortex interaction from the point of view of the 
IIA NS 5-brane worldvolume theory. 

\subsec{Fractional D0-branes lifted to M-theory.}
\subseclab{\mthSS}

From the M-theory point of view fractional D0-branes 
have a very nice interpretation:
Since a D0-brane in $N_c$ D4-branes is momentum 
in the M5-brane in the eleventh direction. 
If we think of the $N_c$ M5-branes as a single M5-brane 
wrapped $N_c$ times around the eleventh direction, then a 
wavefunction should be periodic in $\k\r$ rather than $\r$. 
So rather than being a Kaluza-Klein mode with mass ${1\over\r}$
a fractional D0-brane is a Kaluza-Klein mode with mass
${1\over\k\r}$. 
Related observations have been made for D-branes \juan\aki\gavin\rajesh.
This point of view also explains why the fractional D0-branes appear 
as point-like instantons in the D4-brane worldvolume. D0-brane on a 
single D4-brane have no size corresponding to the fact that the 
single M5-brane is trivial. 
A single M5-brane 
wound $\k$ times also has a sector where it looks trivial, indicating that 
the instantons with charge ${1\over\k}$ also have no size in the 
$U(\k)$ gauge theory. Of course, the multiply wound 
single M5-brane is really non-trivial since different parts of the M5-brane 
can interact with each other. This is seen in the D0-D4 system from the 
fact that mulitply fractional instantons can join together to form 
instantons that can have non-zero size.

\newsec{Supersymmetric QCD.}
\seclab{\sqcdNS}

\subsec{SQCD with $N_f < N_c -1$.}
\subseclab{\nfLTncSS}

Now we consider $SU(N_c)$ gauge theory with 
$N_f$ flavors $Q_i$ in the fundamental representation and 
$N_f$ flavors $\tilde Q_i$ in the anti-fundamental representation.
There are many reviews of this subject \neqoneREVIEW.
In the region where $N_f < N_c$,
this theory is known to have a dynamically generated superpotential 
that has the form
\eqn\dyn{W = C\left( {\Lambda^{b_0}\over \det (M)}\right) 
^{1\over N_c - N_f}}
where $b_0 = 3N_c - N_f$ and $C$ is a numerical constant.
For the case, $N_f < N_c -1$, we can see from 
field theory that fractional instantons will play a role
since there are not enough zero modes for an instanton to generate 
equation \dyn. Expectation values for the fundamental fields $N_f$
break the $SU(\k)$ gauge group to $SU(N_c - N_f)$ \SYM. In the unbroken 
gauge group, we can have gaugino condensation as was 
explained in section \gauginoNS. The low energy superpotential 
\eqn\Wlow{ W_L = \Lambda^3}
is related to \dyn\ by the matching relation 
\eqn\match{ \Lambda^{b_{0_L}} = {\Lambda^{b_0} \over \det (M)}.}

To see SQCD in string theory,
we will use the 
same construction as in the case for \SYM involving 
D4-branes suspended between NS 5-branes and NS' 5-branes, only now we will
add D6-branes extending in the directions
$(x^0,x^1,x^2,x^3,x^7,x^8,x^9)$. The D4-D6 strings provide fields 
in the  fundamental 
representation. The D4-branes can now break along the D6-branes and 
move in the $x^7,x^8,x^9$ directions. The relative position 
of the pieces 
of D4-brane correspond to 
expectation values of the meson fields $M_{i,j} = Q_i\tilde Q_j$.
Because the s-configuration rule 
allows for only a single D4-brane to stretch between a D6-brane and a 
NS 5-brane \hw, 
there will be always be $N_c - N_f$ D4 branes left unbroken making it
possible for a D0 brane to fractionate. Fractional 
D0-branes induce gaugino condensation in the 
worldvolume theory in the same way as discussed 
in section \gauginoNS.

\subsec{SQCD with $N_f = N_c -1$.}
\subseclab{\ncMINUSoneSS}

\ifig\inst{This is the 't Hooft vertex. The dashed lines 
represent gaugino zero modes. The dotted lines represent quark 
zero modes, and the solid lines represent squark expectation values.
This diagram leads to the Afflect-Dine-Seiberg superpotential giving 
a mass to the quarks at finite squark vacuum expectation value.
This is also the diagram of a D0-brane with D4 and D6 strings ending on it 
in the brane construction of SQCD with $N_f = N_c - 1$ flavors.}
{\epsfxsize2.0in\epsfbox{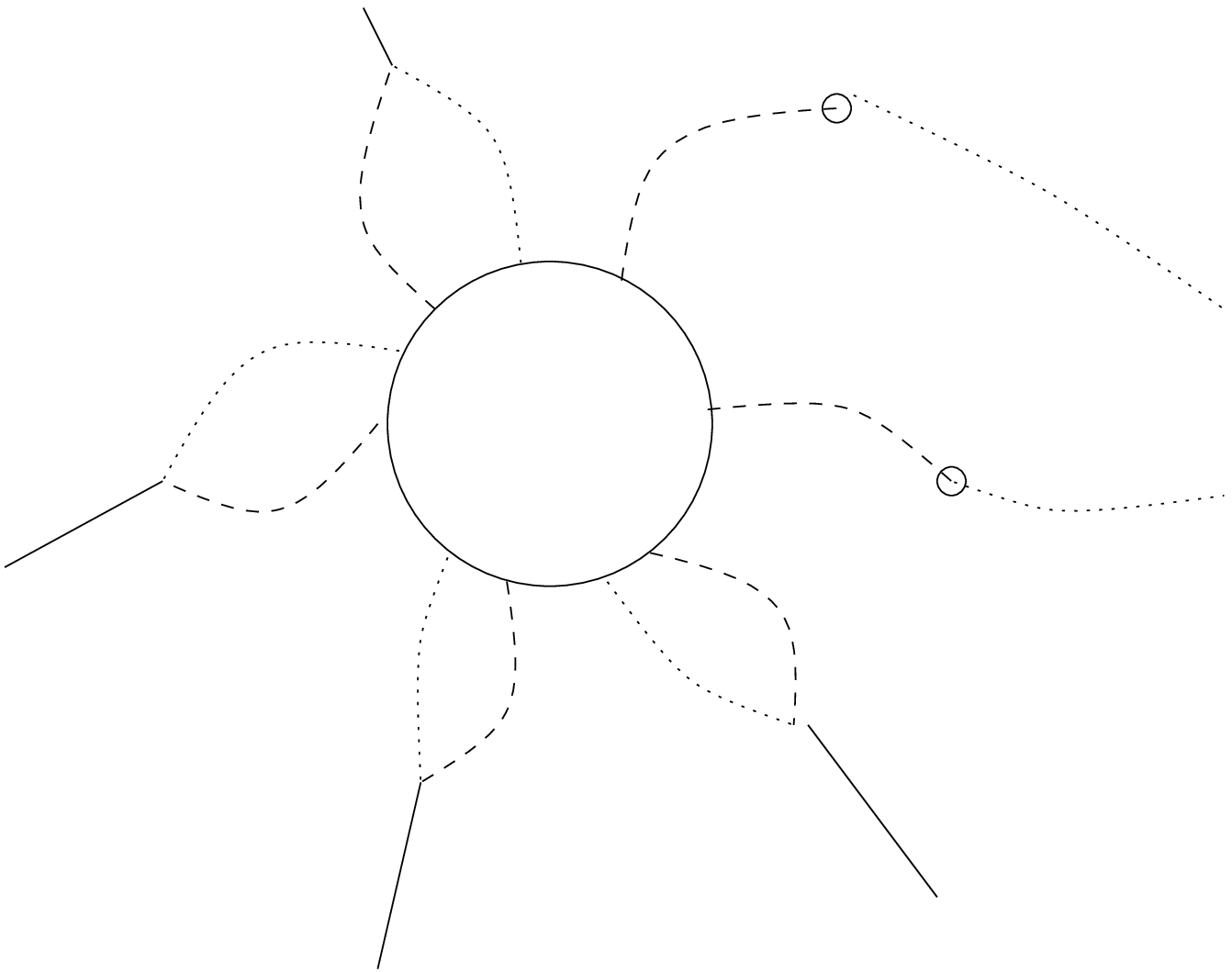}}

For the case, $N_f = N_c -1$, the superpotential 
can be shown to be generated by instantons.
An instanton in SQCD has $2N_f$ zero modes coming form the 
quarks, $\psi_Q$ (the fermionic components of $Q$), 
and $2N_c$ zero modes coming from the 
gauginos, $\lambda$.
Using the 't Hooft vertex, we attach $2N_f$ quark 
``legs'' and $2N_c$ gaugino legs. Because there is the gauge-invariant 
term in the action 
\eqn\yukawa{{\cal L} = ... + \phi_Q \psi_{\tilde Q} \lambda}
we can use it to tie the \gluino\ legs to the 
quark legs (see \inst ). Here $\phi$ is the scalar part 
of $Q$. 
The correlator then roughly looks like 
\eqn\instVERTEX{<\phi^{2N_f} \psi_Q \psi_{\tilde Q}> \simeq \Lambda^{b_0}.}
We see that as ${\phi\over \Lambda} \rightarrow \infty$, $<\psi \psi> 
\rightarrow 0$. Therefore, there is a supersymmetric vacuum far out 
along the Higgs branch, but everywhere else there is an unstable 
vacuum forcing us to infinity. This is in agreement with the Witten index 
which predicts that SQCD has $N_c$ supersymmetric vacua. 

How can we see such an instanton-generated superpotential 
using brane constructions? 
Because of the s-configuration, 
for $N_f = N_c -1$, there will always be 
one unbroken D4-brane. 
Introducing a D0-brane 
corresponds to  
considering the one instanton sector of the theory, and 
we know from above that there should be a dynamically generated superpotential.
In fact, the strings ending on the D0 brane are very reminiscent of 
the ``legs'' of the t' Hooft vertex. There is only one 
unbroken D4 brane providing D0-D4 strings giving
the fermion zero modes $\psi\psi$ where as the 
other D0-D4 strings are stretched and contribute a 
scalar vacuum expectation value $\phi^{2N_f}$.
In the presence of a 
D0 brane, the  
vortex-vortex interactions on the NS 5-brane worldvolume
do not cancel and there is a repulsive force
between the D4-branes driving the 
D4-brane pieces off to infinity along the D6-branes in the 
$x^8,x^9$ direction. This is 
exactly what we expect from the 
form of the dynamically generated superpotential
\dyn. 

\subsec{SQCD with $N_f = N_c$.}
\subseclab{\ncEQnfSS}

Now lets look at the SQCD case where $N_f = N_c$.
Clearly, the superpotential term \dyn\ doesn't make any sense
since $N_c - N_f = 0$. 
The instanton calculation shows that there is 
no dynamically generated superpotential but rather 
a quantum mechanical 
modification to the classical constraint on the 
Higgs branch which is 
\eqn\blowup{\det (M) - B\tilde B = \Lambda^{2N_f}}
\foot {For $U(N_c)$ theories, this would just be $\det (M) = \Lambda^{2N_f}$.}.
We can see how instantons generate \blowup\ by
adding $2N_f$ quark legs and \gluino\ legs to the 't Hooft vertex.
Because of \yukawa, we can join the two types of 
fermionic legs together giving $2N_f$ scalars. 
This then gives \blowup.

How can we understand the quantum modified constraint \blowup\ from 
the brane configurations? Because there are equal numbers of 
D6 brane as there are D4-branes, all of the D4-branes can break on the 
D6 branes. Therefore, generically there are no unbroken D4-branes at the origin 
of moduli space where the D0 brane is. The heavy D4 strings ending on the 
D0 brane correspond to $2N_f$ scalar fields with vacuum 
expectation values. Because generically 
there is never an unbroken D4-brane, 
supersymmetry is generically locally broken to eight supercharges, 
and therefore we 
expect the monopole-monopole interactions to cancel except for 
the special situation where pieces of a broken D4-brane join 
together at the origin. A D4-brane coincident with a D0-brane at the 
origin generates a force 
driving the other broken D4-branes off to infinity. This is 
just the type of behavior we expect from \blowup: As 
one of the meson's expectation value becomes small, 
the other's must become large.

The D0 brane in the brane configuration also 
explain why chiral symmetry breaks in this 
case. It is not possible for the 
D4-branes to become coincident at the 
origin. Therefore, it is not possible for 
the D6-branes to split on the NS' 5-brane 
enhancing the global symmetry group from $SU(N_f)$ to $SU(N_f)_L 
\times SU(N_f)_R$ \brodieami.

\subsec{SQCD with $N_f > N_c$.}
\subseclab{\nfGTncSS}

For SQCD with $N_f > N_c$, there is no quantum modification to 
the moduli space. It is in these cases that there is a 
dual description of the physics at the origin of moduli space.
This is well described in the brane configuration by exchanging 
the NS' with the NS in the $x^6$ direction, a continuation past 
infinite coupling. Since the D4-branes are repelled from the 
origin by the presence of the D0-branes, this suggests that 
the lowest energy configuration may be the dual brane 
configuration.

\newsec{The Intriligator-Thomas-Izawa-Yanagida model.}
\seclab{\itiyNS}

Now that we understand instanton dynamics in terms of 
branes, it is natural to look at another interesting example
considered in \kenscott\iy.
M-theory constructions were also considered in 
\berkeleyIT.
This is SQCD where $N_f = N_c$, but now also have scalar field $S$
that couple to the fundamental fields
\foot{Here we consider the $U(N_c)$ version of the theory which 
has no baryons.} 
\eqn\scalar{W = SQ\tilde Q.}
A supersymmetric minimum demands that 
$Q\tilde Q = 0$, on the other hand we know that 
there is the quantum modified constraint in this case 
that forces $\det M$ to be non-zero \blowup.
Therefore, quantum mechanically, 
there is no supersymmetric minimum.
Although this theory is a non-chiral theory,
supersymmetry breaking is not violated since
the fundamental flavors cannot be given a mass.

We can now consider a brane construction of this theory.
We consider adding D6 branes that lie to the left of the 
NS' 5-brane. The NS 5-brane lies to the right of the 
NS'. $N_f$ D4-branes 
extend between the D6-branes and the NS' 5-brane and $N_c$ D4-branes 
extend between the NS' and the NS 5-brane. An 
expectation value for the scalar component of the 
singlet field 
$S$ corresponds to moving the $N_f$ D4-branes in the $x^8,x^9$ 
directions. An expectation value for $S$ gives a mass to the 
fundamental fields, and we recover the 
SYM case. As in the field theory, the 
fundamentals scalars cannot acquire vacuum expectation values 
since 
breaking the D4-branes on the D6-branes 
would produce an s-configuration.
Adding a D0 brane again induces a force between the 
D4-branes which would like to move away from each other 
in the $x^8,x^9$ direction. However, the
D4-brane are constrained to lie at the origin by the
configuration of D6, NS', and NS 5-branes.
Because the force between the 
D4-branes does not cancel, there is a 
non-zero vacuum energy, and supersymmetry must break.

\newsec{Instantons in theories with $N=2$ supersymmetry.}
\seclab{\neqtwoNS}

Here we consider $SU(N_c)$ with $N=2$ supersymmetry.
In $N=1$ terminology, this is like having 
a single adjoint chiral field. Therefore there 
are $2N_c$ zero modes coming from the \gluinos\ and 
$2N_c$ zero modes coming from the fermion 
component of the adjoint chiral multiplet.
In the 't Hooft vertex, we can again join the legs of the two types 
of fermion zero modes together giving
\eqn\uplane{(\det \phi)^2 = \Lambda^{2\k}.}
As in the SQCD case with equal numbers of colors and flavors,
the origin of moduli space appears to have been removed.
In the $SU(2)$ Seiberg-Witten solution there is a monopole 
at the point $u = \Lambda^2$ and a dyon at the point 
$u = - \Lambda^2$. In the dual coordinates, this is shown in the 
low energy effective superpotential valid near the origin
\eqn\Wdual{W_d = (u + \Lambda^2)q\tilde q.}

We can see in the brane diagram why this is so.
The $N=2$ brane construction has $N_c$ D4-brane suspended between 
two NS 5-branes. The D4-branes can move in the 
$x^4,x^5$ direction: This is the Coulomb branch.
By putting D0-branes at the origin, we see that this case is 
very reminiscent of SQCD with $N_f = N_c$. If a D4 brane is placed on 
top of 
the D0-brane, it breaks more supersymmetry allowing the 
D4-brane to exert a force on its neighboring 
D4-branes, repelling them.
We also see why in this case, as in the cases $N_f \geq N_c - 1$
that fractional instantons don't play a role. There is never 
a point on the moduli space where unbroken D4 branes are 
coincident. Therefore, the D0-branes never have the opportunity 
to split. This explains why fractional branes never played a role in 
the Seiberg-Witten solution.
We can also consider what happens when we put a D2-brane 
in the directions $x^0,x^4,x^6$. The D2-brane is a 
monopole in the 3+1 worldvolume theory. Classically, the mass of the 
monopole is given by the area of the separation between the 
D4-brane and NS 5-brane. According to 
\Wdual\ including the D0-brane we find that 
the mass of the monopole must change by an amount 
proportional to $\Lambda^2$. By adding momentum to the 
membrane along the eleventh direction we have increased 
the membrane mass
as expected.

\newsec{$N=1$ $Sp(N_c)$ with an antisymmetric tensor field.}
\seclab{\spNS}

An interesting model to consider is $N=1$ $Sp(N_c)$ 
gauge theory with an traceless antisymmetric field $X$. Expectation 
values for the $X = diag (a_1,a_2, ..., a_{\k})$ break $Sp(N_c)$ down to 
$Sp(1)^N_c$ where $\sum_i a_i = 0$. 
Because $Sp(1) \simeq SU(2)$, there will be gaugino 
condensation in each of the $Sp(1)$ giving rise to a superpotential
\eqn\Spsup{W = \pm\Lambda_1^3 
\pm\Lambda_2^3 + \cdots \pm\Lambda_{\k}^3}
where $\Lambda_i$ is the dynamically generated 
scale of the $i$-th 
$Sp(1)$ theory. It is possible to have all of the terms in 
\Spsup\ cancel such that $W = 0$. The scales $\Lambda_i$ 
are related to the size of the vacuum expectation 
values $a_i$ and the scale of the unbroken $Sp(N_c)$ 
$\Lambda$ by the 
matching relation 
\eqn\SpMATCH{\Lambda^{N_c+2} = \Lambda_k^3 \prod_{i \neq j \neq k} 
(a_i - a_j).}
Plugging these relations into 
\Spsup\ we find 
\eqn\SpHIGH{W = \Lambda^{N_c + 2}
\left( {\pm (a_1 - a_2) \pm (a_2 - a_3)  ... \pm (a_{\k -1} - a_{\k})
\over \prod_{i\neq j} (a_i-a_j)}\right) .}
Chosing the signs correctly in \SpHIGH\ we can 
have a moduli space of 
the $Sp(N_c)$ theory where there is no superpotential \cho\skiba\bci.
There will be another branch where the terms in 
\SpHIGH\ do not cancel and there is a dynamically generated superpotential.

The brane construction for $Sp(N_c)$ is the same as 
the configuration for 
$N=2$ $SU(N_c)$ SYM with the introduction of 
an orientifold 6-plane. 
The O6-plane projects out the symmetric part 
of the adjoint leaving an antisymmetric field.
Motion of the D4-branes in the $x^4,x^5$ direction 
corresponds to giving a vacuum expectation value to 
the antisymmetric field breaking $Sp(N_c)$ to $Sp(1)^{\k}$. 
Inserting D0-branes into each 
of the $\k$ 
D4-branes generates a mass gap as discussed in section \gauginoNS\ above.
However it is possible to arrange all the 
D0-branes such that the vacuum energy in the $Sp(\k)$ cancels.
Therefore, the effects of the D0-branes can cancel leaving a 
unlifted moduli space.
As an example, this cancellation of D0-branes is easiest to 
see in the case of $Sp(2)$ broken to $Sp(1) \times Sp(1)$. 
Here length of the D0-branes in the D4-branes is the same in the 
$x^6$ direction so the $x^{11}$ momentum through a surface $L_6$ is 
the same for each of the two D0-D4 bound state. 
The mass gap generated by the two D0-D4 systems is the same magnitude.
We then just take a D0-brane in one of the D4-branes to be 
momentum in $-\x$ and the 
D0-brane in the other D4-brane to be momentum in $-\x$. The D0-brane 
effects cancel. However what about $Sp(3)$?
How can a combination of three D0-branes and anti-D0-branes add to zero?
The solution is that the NS 5-branes bend \bending. The bending 
make the Euclidean D0-branes longer in the $x^6$ direction causing the 
mass gap energy per unit volume to decrease. 
The effect of the D0-brane goes as 
$\Lambda^{b_0} \simeq e^{-L_6\over \r}$.
Therefore, the two D0-D4 systems with larger $L_6$ will contribute less 
to the mass gap energy 
while the D4-brane in the middle with the shortest $L_6$ contributes 
the most. The mass gap energies in the three vacua can cancel leaving $W = 0$.
The other branch of the moduli space are given when the 
D0-brane effects do not cancel. In the case of no cancellation, 
there is a repulsive force between 
the D4-branes pushing them to infinity in the $x^4,x^5$ direction.

\newsec{Confinement due to Merons.}
\seclab{\confineNS}
\subsec{Merons}
\subseclab{\meronsSS}

Having explored how instantons generate superpotentials and modify
the classical moduli space, let's see what effect instantons have on 
confinement in $N=1$ \SYM. Although $N=1$ \SYM\ is known 
to confine, it is not known what the mechanism is. 
Here we will argue that 
confinement of electric flux is due to 
point-like fractional instantons, merons.
Merons 
have been shown to be crucial for 
understanding confinement in non-supersymmetric Yang-Mills
\cdg. By placing merons into the interior 
of a Wilson loop 
it was shown that the energy becomes proportional to the area of the loop, 
indicating the
onset of confinement. 
Placing whole instantons into the 
Wilson loop was not enough for inducing confinement. 

\subsec{4+1 SYM with 16 supercharges.}
\subseclab{\sixteenSS}

Let's consider a D2-brane extending in the $x^0,x^1,x^2$ direction and 
a D4-brane extending in the $x^0, x^1,x^2,x^3,x^4$ directions.
When we lift such a configuration to M-theory, the 
D2-brane remains a membrane and the D4-brane becomes a 
5-brane. We therefore have a 
M2 brane inside a M5 brane. Upon returning to IIA, 
if the D2-brane in the D4-brane is 
to remain a BPS object, it must become unconfined magnetic flux, 
$B_{3,4}$ (assuming that $x^1,x^2$ directions are compact). 
We can see this from the 
term in the D4-brane worldvolume action \douglas 
\eqn\flux{{\cal L} = C^{(3)}_{R-R} \wedge F^{(2)}+ ....}
From the eleven-dimensional point of view, the M2 brane is 
H-flux in the M5-brane $H_{0,1,2}$ since 
\eqn\Mflux{{\cal L} = C^{(3)} \wedge H^{(3)} + ....}
where $C^{(3)}$ is the RR field that couples to 
the D2-brane.

By similar reasoning 
to the D2-D4 system, 
the D0-brane must become magnetic flux inside of the D2-brane. In fact, the 
D0-D2 system is T-dual to the D2-D4 system. Non-zero magnetic 
flux in the 2+1 D2-brane worldvolume theory
\eqn\mag{[D_1,D_2] = F_{1,2} \neq 0}
maps to D-term conditions on the 
scalar fields of the 0+1 gauge theory 
in the T-dual D0-brane worldvolume theory,
$[X_1,X_2] = A/\N$ where $A$ is the area of the 
torus and $\N$ is the number of D0-branes \watiREVIEW. 
Therefore, magnetic flux in a D2-brane is T-dual to a 
membrane built from D0-branes on a non-commuting torus. 
In the large D0-brane limit we recover commuting space-time.
Now consider D0-branes inside D4-branes.
If we have $\k > 1$ D4-branes, it is not true that 
D0-branes must be localized since the corresponding instantons 
can have size. This is the reason that merons, 
fractionally charged point-like instantons, are 
so appealing for confinement.
Building a membrane out of fractional D0-branes in 
$N_c$ D4-branes,
it follows that the membrane  
is a localized vortex in the 4+1 SYM.

We can also consider what happens if we have Euclidean 
D0-branes in the D2-brane worldvolume. Euclidean D0-branes induce 
{\it electric} flux along the D2-brane worldvolume $F_{0,1} \neq 0$.
Following the above prescription, we T-dualize the D0-D2 configuration 
twice along $x^0,x^1$. We now have a membrane built from Euclidean 
D0-branes where $[X_0,X_1] \neq 0$ implying that 
space and time generically do not commute.
In the large instanton limit we can recover commuting space-time.
Again embedding this configuration of D0-branes into a D4-brane, 
we would expect the D2-brane to appear as a confined flux.
It is easy to see by U-duality that we can also build a fundamental string 
out of Euclidean D0-branes.

A way of seeing why the lowest energy configuration 
of the D2-brane inside the D4-brane is 
spread out in the absence of D0-branes and confined 
in the presense of D0-branes is to consider a U-dual 
configuration of perpendicular D1-branes and fundamental strings.
Consider non-Euclidean D0 branes in $(x^0)$, D2-branes in $(x^0,x^1,x^2)$, 
and D4-branes in $(x^0,x^1,x^2,x^3,x^4)$. 
An operation $T_1 T_2 S T_3$ 
on this configuration should bring us to a 
D1-D1-NS1 system where one set of $\k$ 
D1-strings are extended along $x^0,x^4$,
the other set of D1 strings and NS strings are extended along $x^0,x^3$.
We take $x^4$ to be the vertical direction and $x^3$ to be 
horizontal.
In the absence of fundamental strings along $x^3$, the lowest 
energy configuration of the D1-branes
along $x^3$ and $x^4$ is to be tilted  
diagonally in $x^3,x^4$.
This diagonal configuration of a D1-brane 
is clearly BPS since it is just a 
D1-brane in a rotated coordinate system.
Because of the tilt, when we T-dualize along 
either the vertical or horizontal directions, 
we get magnetic flux 
on a D2-brane \watiREVIEW. If we continue to 
U-dualize, the tilted D1 brane can eventually be related to  
magnetic flux in a D4-brane. 
However, going back to the D1-D1 system, 
if we allow for NS strings along $x^3$ 
the D1-branes, the NS-strings and the D-strings form a bound 
state, and it is easy to 
see that the D1-branes lowest energy configuration 
will not be the tilted one, but the configuration of 
vertical D1-branes and horizontal D1-branes.
U-dualizing, this configuration transforms it 
into a membrane of confined magnetic
flux in a D4-brane.
Notice however that the configuration can be BPS only 
when the flux has spread out. 
A D2 brane in a D4-brane that does not spread out
is not BPS as can be seen easily from the supersymmetry relations. 
This is unfortunate but in agreement with the findings 
of MQCD \qcd.

\subsec{Wilson loops.}
\subseclab{\wilsonSS}

\ifig\confine{In the figure on the left we have a D4-branes with a 
fundamental IIA string inside it. The lowest energy configuration 
is for the string to spread out within the D4-brane. From the point of 
view of the D4-brane worldvolume the string looks like 
unconfined electric flux. In the figure on the right, we have put 
Euclidean D0 branes into the worldvolume of the D4-brane. Because the 
string ends on the Euclidean D0-brane, 
the flux must flow through it. One can show that
the lowest energy configuration is now for the string to be confined 
electric flux in the D4-brane worldvolume.}
{\epsfxsize4.0in\epsfbox{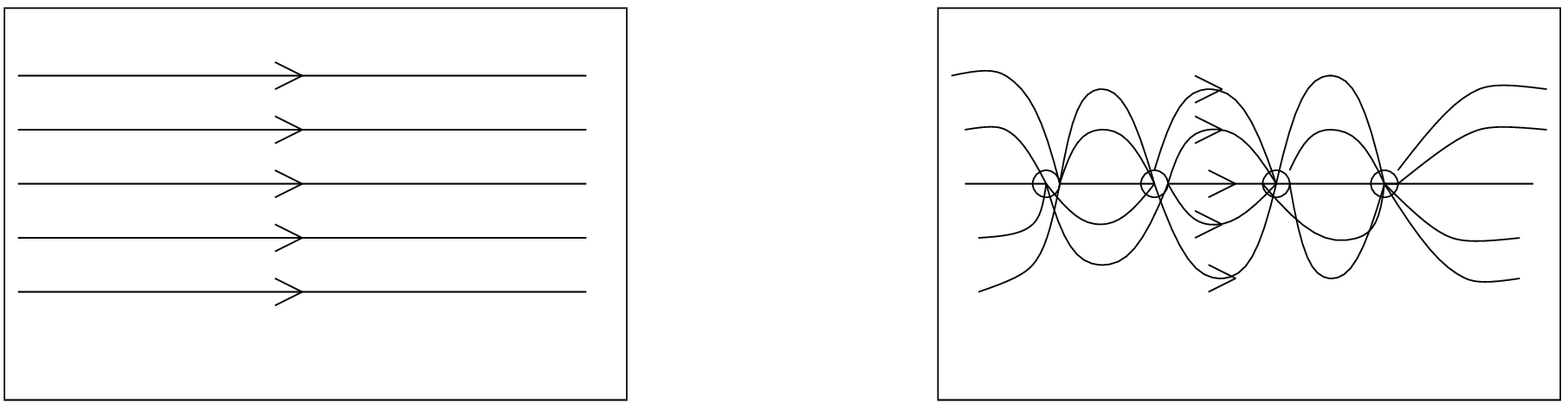}}
A typical method for demonstrating confinement is to use a Wilson loop. 
The Wilson loop is created by taking a heavy quark and anti-quark, separating 
them a distance $y$ for a time $t$ and then having the 
quarks annihilate. If the energy of the loop as compared to 
the vacuum goes as the area, $yt$, then the theory is confining. 

To create the Wilson loop in the D-branes, we can take two D4-branes separated 
by a distance $L$. Let us consider two strings joining the 
two D4-branes. The are oppositely oriented and separated a distance 
$y$. Typically, when a fundamental string ends on a D-brane, it appears 
in the worldvolume theory as a charged particle (quark) that couples to 
the worldvolume gauge field. Gluons are string oscillations 
on the brane that move away from the point at which the oriented string 
ends. Classically, there is a Coulomb-like force that 
attracts the quark and anti-quark. The attractive force will 
cause the strings to move together joining and then radiating 
away into the bulk as a closed string. The Wilson loop clearly does not 
obey an area law. 

What happens if we include Euclidean 
D0-branes, instantons in 
the worldvolume theory. 
As we saw above, using point-like Euclidean 
D0-branes we can build a membrane or 
fundamental 
string from the D0-branes.
A string built from Euclidean D0-branes is confined electric 
flux in the D4-brane worldvolume.
The charge from the string 
is then passed from D0-brane to D0-brane and then to the oppositely oriented 
string forming a flux tube. The quarks are never seen in 
the worldvolume theory. They are truely confined by the instantons. There 
is an attractive force between the confined quarks that is linear and 
depends on the mass of the D0-branes. The string and D0-branes 
annihilate into the bulk after a time $t$.

\subsec{$N=1$ SYM.}
\subseclab{\symSS}

Now we return to $N=1$ SYM.
At strong coupling $N=1$ SYM is known 
to confine. Using the brane construction described 
in section \gauginoNS,
we use Euclidean D0-branes along $x^6$
to build a IIA string along $x^0,x^1$ in the worldvolume 
of the D4-brane.
Having the 5-branes dimensionally reduces the 
D4-brane in the $x^6$ direction makes 
the theory 3+1 dimensional, and therefore the IIA strings really have the 
interpretation as vortices.
We can also have a string in the 3+1 worldvolume 
by building a D2-brane along $x^0,x^1,x^6$ from D0-branes 
along $x^6$. This string should be 
confined magnetic flux.
There are also two types of instantons in 3+1, electric and magnetic.
An electric instanton is the one we have been discussing 
which has action ${L_6\over\r}$ and is momentum $T^{6,11}$ in the 
$(0,2)$ theory. A magnetic instantons has action ${\r \over L_6}$ 
that is momentum $T^{11,6}$. It is 
interesting that the symmetry between electric and magnetic 
instantons is related to the fact that the energy momentum tensor 
is symmetric.

Because the theory of $\k$ D4-branes is $SU(\k)$ SYM with only adjoint matter 
we can think of it as $SU(\k)/Z_{\k}$ since the adjoint is invariant under 
the center of $SU(\k)$. $\pi_1(SU(\k)/Z_{\k}) = Z_{\k}$ and so 
this theory can have $Z_{\k}$ types of topologically stable vortices.
This is 
easy to see from their close relationship to monopoles.  
We would like to associate each of the $\k$ vortices 
with $\k$ fractional instantons. Once the 
$\k$ vortices join together, 
$\k$ fraction instantons have joined together also.
The D0 branes can leave the D4-brane 
as a graviton and so can take the fundamental string with it. 
In this way, the $\k$ types of 
strings can annihilate.

\newsec{Acknowledgements}
I would like to thank S. Ramgoolam, A. Hanany, A. Hashimoto, 
K. Intriligator, Y. Oz, J. de Boer, 
K. Hori, M. Strassler, W. Taylor, P. Pouliot, and E. Witten 
for helpful 
discussions. I would also like to thank the physics departments 
of University of California at Berkeley, Stanford Linear Accelerator Center, 
California Institute of Technology, and 
University of California San Deigo for their hospitality 
during the undertaking of this work. The work is supported by 
Dept. of Energy under Grant DOE-FG02-91ER40671.

\listrefs

\end